# The Non-Axisymmetric Influence: Radius and Angle-Dependent Trends in a Barred Galaxy

Carrie Filion,[1][*] Rachel L. McClure,[2] Martin D. Weinberg,[3] Elena D'Onghia,[2] Kathryne J. Daniel[4]

[1]*William H. Miller III Department of Physics & Astronomy, The Johns Hopkins University, Baltimore, MD 21218*
[2]*Department of Astronomy, University of Wisconsin–Madison, Madison, WI,53706*
[3]*Department of Astronomy, University of Massachusetts at Amherst, Amherst, MA 01003, USA*
[4]*Department of Astronomy & Steward Observatory, University of Arizona, Tucson, AZ, 85721*

**ABSTRACT**

Many disc galaxies host galactic bars, which exert time-dependent, non-axisymmetric forces that can alter the orbits of stars. There should be both angle and radius-dependence in the resulting radial rearrangement of stars ('radial mixing') due to a bar; we present here novel results and trends through analysis of the joint impact of these factors. We use an N-body simulation to investigate the changes in the radial locations of star particles in a disc after a bar forms by quantifying the change in orbital radii in a series of annuli at different times post bar-formation. We find that the bar induces both azimuth angle- and radius-dependent trends in the median distance that stars have travelled to enter a given annulus. Angle-dependent trends are present at all radii we consider, and the radius-dependent trends roughly divide the disc into three 'zones'. In the inner zone, stars generally originated at larger radii and their orbits evolved inwards. In the outer zone, stars likely originated at smaller radii and their orbits evolved outwards. In the intermediate zone, there is no net inwards or outwards evolution of orbits. We adopt a simple toy model of a radius-dependent initial metallicity gradient and discuss recent observational evidence for angle-dependent stellar metallicity variations in the Milky Way in the context of this model. We briefly comment on the possibility of using observed angle-dependent metallicity trends to learn about the initial metallicity gradient(s) and the radial rearrangement that occurred in the disc.

## 1 INTRODUCTION

In the past half-century, significant effort has gone into defining and understanding the dynamical processes through which stars in disc galaxies can move away from their birth radii. A wide variety of dynamical processes have been proposed and explored, each with varying degrees of clear applicability in the Milky Way. The Milky Way is host to a variety of structures that cause time-dependent gravitational forces and drive long-term orbital evolution. In this analysis, we examine effects from one of the strongest perturbations in the Milky Way – the Galactic bar (see e.g. Blitz & Spergel 1991, Weinberg 1992).

Standard Hamiltonian perturbation theory predicts that a central bar can radially redistribute disc orbits (e.g. Binney & Tremaine 2008), changing both the radial and angular distribution of stars. In a disc with a radius-dependent initial metallicity gradient, azimuth angle-dependent metallicity variations would be an observational result of these radial changes. It is in this context that trends in angle dependence in disc galaxies are typically explored (e.g. Di Matteo et al. 2013, see also the test particle simulation presented in Wheeler et al. 2022 and the more spiral-focused analyses of Grand et al. 2016, Khoperskov et al. 2018, Fragkoudi et al. 2018).

Azimuthal metallicity variations can also be induced from external perturbations. Angle-dependent metallicity variations have been observed in the gas phase of close galaxy pairs with integral field unit spectroscopy (Hwang et al. 2019), and are predicted to arise in stellar populations from galactic mergers (e.g. the Sagittarius-like merger presented in Carr et al. 2022). In the Milky Way, it is likely that both internal non-axisymmetric structures (such as the bar) and external perturbations (such as Sagittarius) are altering the orbits of stars in the disc. It is important to understand the role of each, and in this analysis we focus on the radius and angle-dependent trends that can arise from a bar alone.

The formation and evolution of a bar alters the orbits of stars in a disc in a few key ways. The first is the exchange of angular momentum between stars and the bar as the bar is forming and growing. The bar 'steals' angular momentum from stars in the bar region and transports angular momentum outwards (see e.g. Lynden-Bell & Kalnajs 1972, and more recent discussion in Petersen et al. 2019). The second is through resonant interactions and the bar exerting torques on the stars in the disc, altering the stars' orbital properties. Finally, the existence of a strong non-axisymmetric potential from the bar modifies the radial distribution of stars.

In this analysis we use an N-body simulation to study the evolution of stellar orbits in a time-dependent, barred potential. While N-body simulations omit some of the physical processes that influence galaxy evolution, they are an excellent laboratory within which to study dynamical processes in detail. The simulation that we employ in this analysis is not intended to match the Milky Way, and the results of our analysis are not meant to be detailed predictions for any particular galaxy. The initial conditions for our simulation were chosen such that the resulting galaxy would be bar-dominated and the amplitude of trailing spiral structure would be suppressed. We also sought to avoid vertical instabilities (i.e. bar buckling). Our choice of initial conditions was informed by the goal of our analysis, which is to study the radius- and azimuth angle-dependent trends that can arise in a bar-dominated galaxy due to the time-dependent, evolving non-axisymmetric potential.

[*] E-mail: cfilion@jhu.edu

Unlike many of the earlier analyses of azimuth angle-dependent variations discussed above, we perform our dynamical analysis in terms of 'present-day' and initial (or 'birth') radii, not metallicity, and we tag the initial radius of the stars as their radial location at the time that the bar has reached its maximum amplitude. Framing the analysis in terms of radii (instead of metallicity) allows us to more easily explore how the dynamics from a bar and its evolution alters the orbits of stars within the disc, and what angle- and radius-dependent signature this evolution leaves on the stars in a given annulus.

Further, many studies define the initial (or 'birth') radial position of a star to correspond to its location at some time before the bar forms and then perform their analysis at time(s) post bar-formation. We instead define our initial radial position to correspond to a time *after* the bar has formed. We do this to isolate the effects of bar evolution on the disc, separating these effects from those of the bar formation process. As such, this analysis focuses on 'younger' to 'intermediate age' stars ($\sim 1$ to 4 Gyr), whereas studies such as Di Matteo et al. (2013) investigate 'ancient' stars.

We opt to explore the signatures of bar-driven evolution within annuli at discrete times post bar-formation, such that we can investigate the amplitude of the angle-dependent radial rearrangement as a function of both radial position and time. We make no restrictions on orbital circularity and explore the changes in the (instantaneous) orbital radii of a stars that are the end result of bar-induced dynamical processes. The term 'radial mixing' is occasionally defined to describe any radial change, but to avoid any possible confusion with alternate connotations we explicitly avoid using this term.

We provide the details of our N-body simulation in Section 2. In Section 3 we present the results of our investigation of the radial distances that stars have traversed in terms of the initial and present-day Galactocentric cylindrical radii ($R$) of the stars. We then adopt a simple initial metallicity gradient and discuss angle-dependent metallicity variations in the context of this toy model in Section 4. We discuss broader implications of our dynamical results in Section 6 and conclude in Section 7.

## 2 METHODS

### 2.1 Simulations

We a use self-consistent, collisionless N-body simulation of an isolated, barred galaxy, which was evolved using the basis function expansion code EXP (Weinberg 1999, see Petersen et al. 2022 for a recent description). The initial conditions for this galaxy are nearly identical to the cusped model presented in Table 1 of Petersen et al. (2021) as 'Potential I', and we highlight below where the models differ. We provide the key parameters of the simulation here, and refer to Petersen et al. (2021) for in-depth discussion of the simulation and its initial conditions. A cuspy dark matter profile best matches that found in pure dark matter simulations, motivating our selection. We direct readers to (e.g.) Petersen et al. (2019) for comparison of the angular momentum matter exchange in cored versus cusped profiles, and note that their analysis indicates that there should be little change in the general (radius-dependent) trends presented here for a different choice of initial dark matter profile. We include in the Appendix a plot of the rotation curve of the galaxy at the beginning of the simulation, along with additional figures that further characterise the simulation.

This simulation is in virial units (where $R_{vir} = G = M_{vir} = T = 1$). Converting to physical units can be done with an appropriate choice of scale factors. For example, one could choose the following scaling, which roughly approximates the Milky Way: $R_{vir} = 300$ kpc, $M_{vir} = 1.4 \times 10^{12} M_\odot$, with $T = 2$ Gyr being the resulting unit of time. The galaxy consists of a live dark matter halo of $N_d = 10^7$ particles and a disc of $N_s = 10^6$ particles. The dark matter halo is a modified Navarro-Frenk-White potential (Navarro et al. 1997) that matches the distribution function given for 'Potential I' in Petersen et al. (2021). For this analysis, however, we have set the 'concentration parameter' to equal 20 (instead of 25, as was adopted for 'Potential I') to better match the combined disc and halo rotation curve of the Milky Way (M. Petersen, *priv. comm.*).

The stellar disc of the galaxy has an exponential density distribution in radius and an isothermal-sech[2] distribution in height. The disc mass is equal to $0.025 M_{vir}$, the scale length ($R_d$) is $R_d = 0.01 R_{vir}$, and the scale height ($z_0$, constant throughout the disc) is $z_0 = 0.001 R_{vir}$. The initial value of the Toomre Q parameter (set by the surface density of the disc, the epicyclic frequency, and the radial velocity dispersion) is $\approx 1.4$.

### 2.2 Bar Properties

Structure forms naturally in the disc relatively quickly in the simulation. A strong bar is evident, as well as weaker trailing $m = 2$ arms outside of the bar-dominated region. We compute both the bar strength and length using the relative amplitude of the $m = 2$ Fourier components of the face-on disc, using the following expression,

$$A_2 = \frac{|\sum_j M_j e^{2i\phi_j}|}{\sum_j M_j}, \tag{1}$$

where $M_j$ is the mass of the $j^{th}$ particle and $\phi_j$ is its galactocentric azimuthal ($\phi$) coordinate location (see e.g. Sellwood & Athanassoula 1986, D'Onghia & L. Aguerri 2020, Baba et al. 2022). In computing the bar strength, we compute $A_2$ within a portion of the bar region. Specifically, the summation is performed over all stars within a Galactocentric cylindrical annulus between 0.25 to 0.50 $R_d$ at each time step.

In Figure 1 it is evident that shortly after $t = t_0$, there is a relatively sharp decrease in $A_2$ corresponding to a change in the bar stability. Similar changes in $A_2$ are often associated with the bar buckling instability. The initial conditions of our simulation were carefully chosen to avoid such vertical instabilities, and we verified that the bar does not buckle by both inspecting the edge-on profiles and computing the buckling amplitude (following Equation 6 of Baba et al. 2022, see also references therein) and the asymmetry parameter ($|S_{bar}|$) from Smirnov & Sotnikova 2018 (see Appendix for supporting plots).

We define the time at which the bar has finished forming (hereafter initial time, $t = t_0$) to be when the bar strength reaches its maximum. We present a plot of bar strength over time in Figure 1, where the vertical line indicates $t = t_0$. We analyse the galaxy's radial and angular trends at 0.5, 1.0, 1.5, and 2.0 $T$ after $t_0$ (equivalent to one, two, three, and four gigayear with the scaling given above). Here, $0.5T$ corresponds to approximately three and a half bar rotations as determined by the bar pattern speed at $t = t_0$).

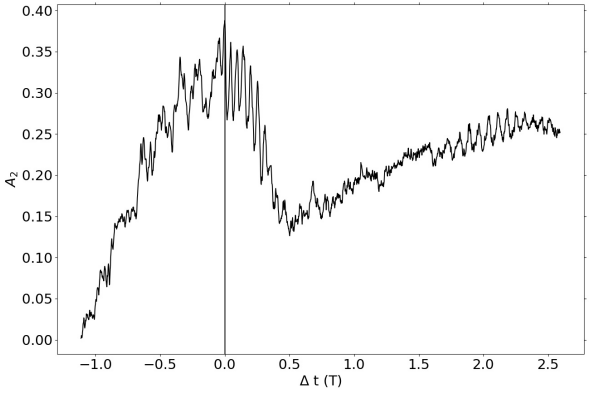

**Figure 1.** The strength of the bar over time, measured within a galactocentric cylindrical annulus between 0.25 to $0.50R_d$. Here, time is given as $\Delta t = t - t_0$, where $t_0$ corresponds to the time where the bar strength is at maximum (also shown as vertical black line).

In computing the bar length ($R_{bar}$), the summation is done over the stars that reside in annular rings of width $\frac{1}{30}R_d$ from $R = \frac{1}{10}R_d$ to $R = 5R_d$ at a given timestep. We identify the maximum of the relative amplitude of the $m = 2$ feature at each time step, and then estimate the length of the bar to be where the relative amplitude drops to $\sim 20\%$ of this maximum value (such that this radius is larger than that where the amplitude is maximum), following e.g. Athanassoula & Misiriotis (2002). We adopt this specific metric as it enables direct comparison to the Milky Way bar, whose length has been estimated in the same way (see e.g. Wegg et al. 2015), but note that it is susceptible to over-estimation due to trailing $m = 2$ structure.

Finally, following Appendix B.3 of Petersen et al. (2021), we compute the approximate location of corotation with the bar ($R_{CR}$) as the radius where the circular frequency and the pattern speed are equal[1]. We present the face-on density plots of the galaxy at each of the selected time steps in Figure 2, where the approximate length of the bar and corotation radii are indicated with a red and black circle, respectively. Note that by $\Delta t = 2.0$ T, under-dense regions along the bar minor axis are evident. Such 'dark gaps' have been interpreted as being associated with corotation and the unstable Lagrange L4 and L5 points (e.g. Buta 2017) or the inner ultra-harmonic resonance (e.g. Krishnarao et al. 2022). We are pursuing a more detailed analysis of resonances, resonant orbits, and their role in radial re-arrangement in this simulation which will be presented in a future manuscript (Filion et al, in prep).

### 2.3 Dynamical Analysis

We select twelve annuli centred on $R = \frac{1}{3}$ through $4R_d$, each of width of $\frac{1}{6}R_d$ (i.e. $\pm\frac{1}{12}R_d$). We investigate the median radial distance that stars in a given annulus traversed to get to their present location as a function of azimuth angle. We refer to this median, angle-dependent radial distance as $\Delta$R. We use the observable *instantaneous* galactocentric cylindrical radius (as opposed to e.g. guiding center radii).

To compute $\Delta$R, we first compute the change in galactocentric radius ($\delta$R, where $\delta$R = $R_t - R_{t_0}$, where $R_{t_0}$ is the initial or 'birth' radius, and negative $\delta$R indicates that the star originated at a larger radius than the present location) for each of the stars that are in a given annulus at each time interval. We divide each annulus into 4°-wide azimuth bins, and $\Delta$R for each bin is then the median value of $\delta$R of that bin. We also compute the angle-averaged mean $\delta$R over all of the stars in each annulus, which we call $\langle\delta R\rangle_\phi$.

We refer to this angle-dependent median distance in a given annulus ($\Delta$R) as the 'bulk' radial change happening in the disc. The bulk trends that we discuss in this analysis cannot describe all of the radial changes happening in a galaxy, as any given star could travel more or less than the median distance. We leave more thorough analysis of individual dynamical families of stars and the mechanisms causing their radial changes to future study (Filion et al, in prep).

### 3 RESULTS

Here, we detail the bulk radial changes that occur in the disc. We restrict this Section to the results of the dynamical analysis, and in Section 4, we make connections to potentially observable signatures.

### 3.1 Trends in the Bulk Radial Changes

In Figure 3 we present $\Delta$R as a function of $\phi$ for all of the annuli at each time interval. For clarity, we include four annuli per column. In each plot, the vertical lines indicate the angle of the bar (measured relative to the positive x-axis) at that time. Angle-dependent, sinusoidal variations in $\Delta$R are seen clearly in most annuli at all times, with the larger radii generally becoming more sinusoidal in time. The strength of alignment of the variations with the bar angle depends on the length of the bar relative to the chosen annulus. Annuli within the extent of the bar are well-aligned with the bar, while those further outside of the bar lag and can be up to ~1 radian offset from the bar.

We present an alternative view of these angle-dependent variations in Figure 4, which shows face-on two-dimensional histograms of the annuli at each time colour-coded by the mean $\delta$R in $\sim\frac{1}{30}R_d$-wide hexagonal bins. Here, density contours of the disc are over plotted in black. The face-on view of the angle-dependent trends are quadrupole-like, and the combination of Figure 3 and 4 reveal that the quadrupole becomes more coherent over time as the bar grows to encompass more of the disc. We note that in each of these plots, we have included all stars in a given annulus regardless of height above the plane, but the results are essentially identical if we restrict the discussion to stars near the plane (e.g. $|z|$ <1 disc scale height).

### 3.2 Implications

At all times, the innermost annuli (left column of Figure 3, where $R \lesssim R_d$ and $R \lesssim R_{bar_0}$, with $R_{bar_0} \sim 1.4R_d$ being the length of the bar at $\Delta t = 0$ T) orbits have evolved inwards and there is a median negative change. The intermediate annuli (middle column, $R \gtrsim R_d$ and $R \gtrsim R_{bar_0}$) are composed of a mix of stars whose orbits have evolved inward and outward, with the outward evolving regions generally aligned with the bar major axis, and inward evolving regions generally aligned with the minor axis. Annuli at radii larger than the bar length are less well-aligned with the bar than those within the bar edge. This worsening alignment is due to the transition to

---

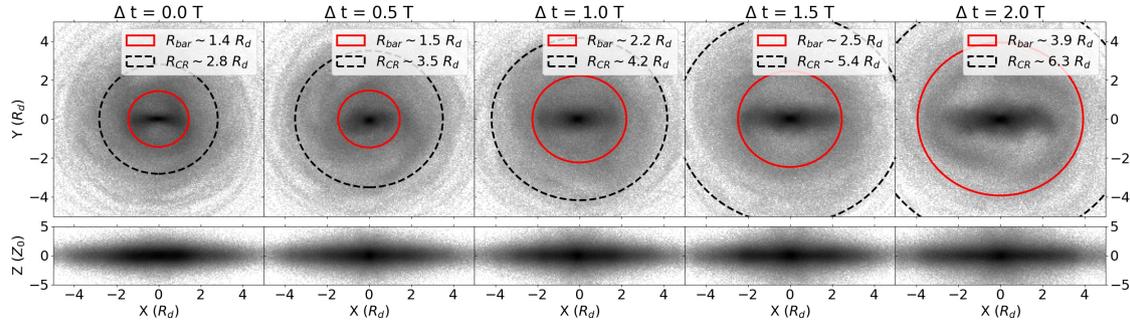

**Figure 2.** The face-on view of the galaxy at $\Delta t = 0$, 1, 2, 3, and 4 gigayear. The colour indicates log-scaled density, with darker grey indicating regions of higher density. Red and black circles indicate the approximate length of the bar and the location of corotation, respectively. Here, corotation is estimated as the radius where the circular frequency and pattern speed are equal. The disc rotates counter-clockwise, and each figure is shown in the frame rotating with the bar such that the bar is aligned with the x-axis.

trailing, spiral-like structure (also an $m = 2$ feature) outside of the bar-dominated region. The outer annuli (right column, $R \gg R_d$ and $R > R_{\mathrm{bar}_0}$) are skewed such that there is a positive median change and orbits have evolved outwards, especially at late times.

The combination of this bulk inward (and outward) evolution with azimuth angle-dependence after bar formation has not been explicitly explored as far as we are aware and should be ubiquitous in galaxies with bar evolution. As we discuss below, in Section 6.2, these effects must be included in semi-analytic models of disc evolution – simple diffusion does not inherently include these bar-driven effects.

The net inwards evolution in the innermost annuli is characteristic of angular momentum loss due to the secular evolution of the bar (see e.g. Petersen et al. 2019), and thus is a generic result for evolving barred galaxies. The effect of secular evolution and dynamical response (i.e. reaction to the non-axisymmetric potential) are coupled. The resulting radial rearrangement can be significant: at $\Delta t = 1.5$ T, for example, stars travelled an angle-averaged radial distance of $\sim \frac{1}{3} R_d$ to enter the annulus centred on $\frac{2}{3} R_d$. That is, stars in that annulus at $\Delta t = 1.5$ T started at radii that were, on average, $\sim 1.5\times$ larger. The average outwards evolution in the outer disc is likely due to the net effect of the bar (and its trailing structure) moving angular momentum outwards, though the radius-dependent decrease in stellar density probably also contributes to this signal (as suggested by Kubryk et al. 2013).

The disc can thus be thought of as having three radial 'zones', and the trends in these zones arise due to the combined effect of secular evolution and dynamical response of orbits. The physical radii that these zones occupy will depend on the properties of the bar, and above we provide the approximate radial ranges of these zones in terms of the $R_d$ and $R_{\mathrm{bar}_0}$ of this simulation. We can further define these zones more precisely using $\langle \delta R \rangle_\phi$. We illustrate these zones and their evolution in Figure 5, which shows $\langle \delta R \rangle_\phi$ and the minimum and maximum $\Delta R$ at each time step alongside the approximate bar length and corotation radii. Here, we define the inner zone to be where $\langle \delta R \rangle_\phi \leq -\frac{1}{12} R_d$, the intermediate zone to be where $-\frac{1}{12} R_d < \langle \delta R \rangle_\phi < \frac{1}{12} R_d$, and the outer zone to be where $\frac{1}{12} R_d \leq \langle \delta R \rangle_\phi$ (note $\frac{1}{12} R_d = 0.25$ kpc adopting the scaling given above). The inner and intermediate zone are both interior to corotation at each time interval. In the inner zone, where orbits have on average evolved inwards, angle-dependent trends are well-aligned with the bar angle and $\Delta R$ and $\langle \delta R \rangle_\phi$ are generally below zero. In the intermediate zone, there is no mean radial evolution and the angle-dependent $\Delta R$ trends are roughly symmetric around zero - i.e. $\langle \delta R \rangle_\phi \sim 0 R_d$. Finally, in the outer zone, orbits generally evolved outwards, $\langle \delta R \rangle_\phi > 0 R_d$, and angle-dependent trends are generally less well-aligned with the bar angle.

We note here that the majority of the annuli in our analysis are within the corotation radius at each time and are within the bar length by late times. Bars in N-body simulations with live halos generally lengthen and slow in time, suggesting that our predictions are most applicable to galaxies with somewhat older bars. We stress that the exact details of our results (such as the radial extents of the zones or the amplitude of $\Delta R$ variations) are dependent on the properties of our simulation and thus are not general predictions for other galaxies. The big-picture trends, however, are more general predictions. Regardless of the location of corotation, for example, there should still be net angular momentum loss of stars in the inner regions of the disk if the bar is growing (see e.g. Petersen et al. 2019) and some net outwards evolution for stars in the outer disk.

### 3.3 Comparison to Pre-Bar Formation Initial Radii

For completeness, we compare our results to those that would be obtained if we instead defined the initial radii to correspond to a time prior to bar formation. We repeat the procedure above, but adopt an initial time of $t = 0$ ($\Delta t \sim -1.1$ T in Figure 1), when there is no structure in the disc. In comparison to the results seen in Figure 3, we find that the shape and phase of the $\phi$ versus $\Delta R$ curves remain largely unchanged, but the annuli centred on radii larger than $\sim R_d$ are generally shifted towards more positive $\langle \delta R \rangle_\phi$ and $\Delta R$ values. The radial locations and extents of each of the zones changes accordingly. For example, at 1 T post bar-formation, the annulus centred on $2R_d$ has $\langle \delta R \rangle_\phi \sim \frac{1}{6} R_d$ if the initial radii of stars is defined at $t = 0$, compared to $\sim 0 R_d$ if the initial radii are defined at $t = t_0$. The general shift outward is likely the same disc expansion mentioned in Di Matteo et al. (2013), and is a sign of the angular momentum rearrangement that must occur in the disc during bar formation.

This exercise also allows us to test the sensitivity of our results to two of our assumptions about how stars are 'born' in galaxies. First, in defining the initial radial locations of stars to correspond to a time after the bar formed, we effectively assumed that the density distribution of newly 'born' stars matches that of the older stars at $t = t_0$. Second, we did not require that stars originate on circular orbits at $t = t_0$. Star formation in barred galaxies is in nature may differ from this extreme, but comparison to the opposite extreme (i.e. the smooth disc at $t_0 = 0$) indicates that our results are largely insensitive

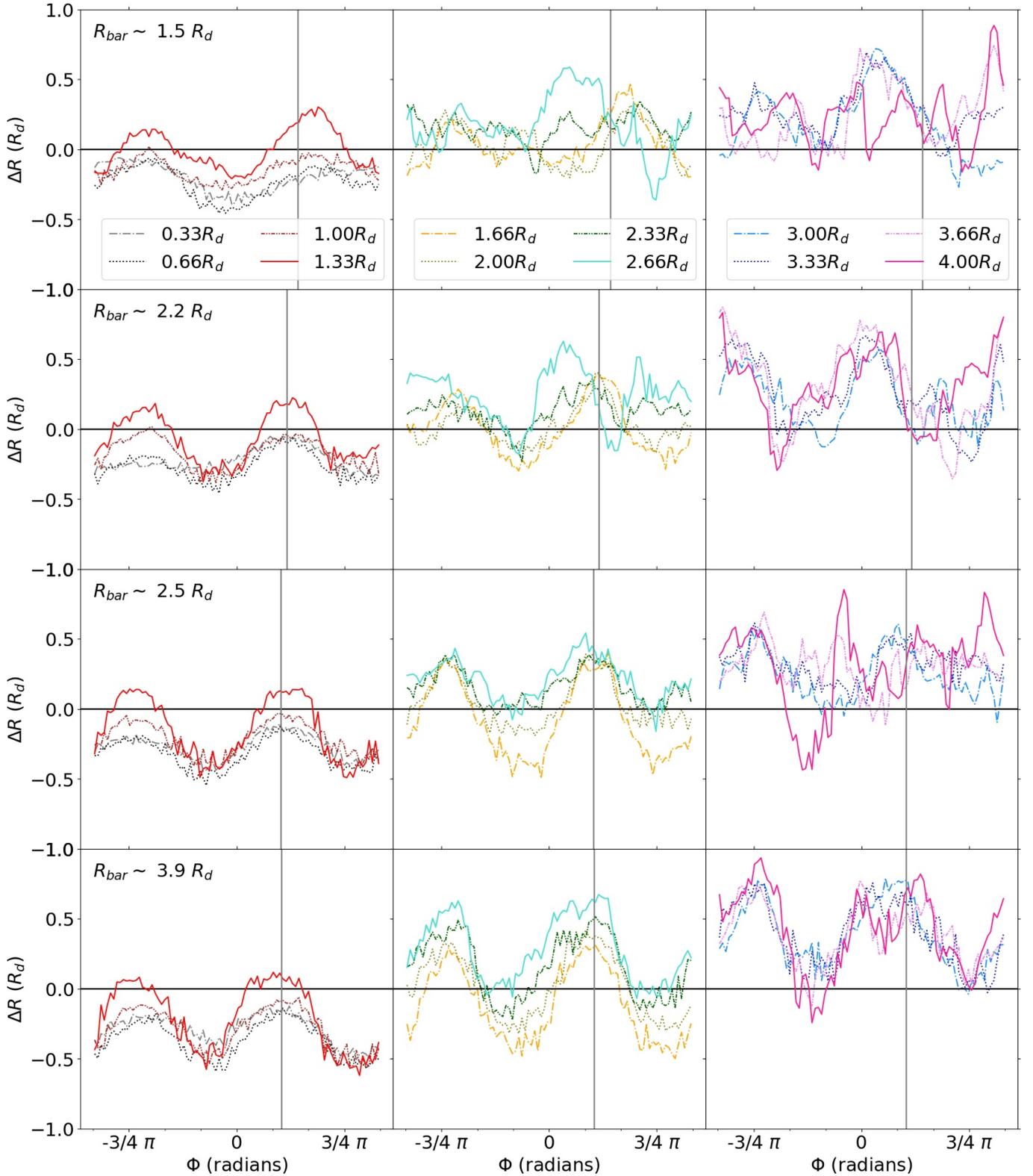

**Figure 3.** The ΔR (i.e. median distance stars have travelled computed in 4° bins) of stars in each of the twelve annuli as a function of angle at the four time intervals considered in this analysis (rows correspond to Δt = 0.5, 1.0, 1.5, 2.0 T from top to bottom). Vertical grey line indicates the angle of the bar at that time. Left column: the four innermost annuli, centred on $\frac{1}{3}$ through $\frac{4}{3}R_d$. Middle column: the four middle annuli, centred on $\frac{5}{3}$ through $\frac{8}{3}R_d$. Right column: the four outermost annuli, centred on 3 through $4R_d$. The center of each annulus is indicated with the colour and line style pairing given in the legend. The three innermost annuli (left column) are entirely below the x-axis at all times, indicating that orbits that have evolved inward in these annuli. The four outermost annuli (right column) are predominantly above the x-axis, indicating that orbits that have evolved outward in these annuli. The trends in the six outermost annuli (middle and right column) become more sinusoidal and aligned with the bar angle at later times (lower rows).

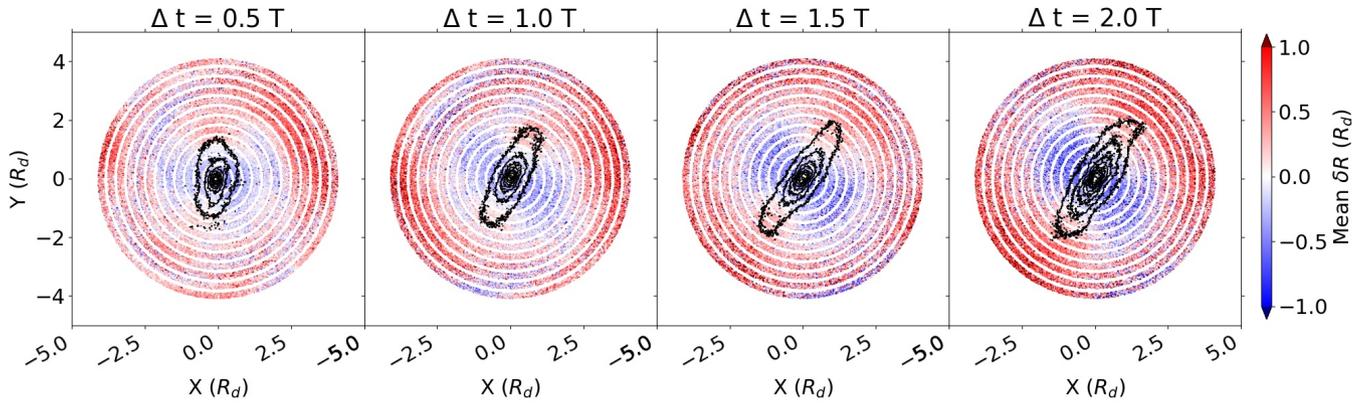

**Figure 4.** Two-dimensional histograms in $X$ and $Y$ of the twelve annuli at each time, colour-coded by the mean $\delta R$ in each $\sim \frac{1}{30} R_d$-wide hexagonal bin. Over-plotted in black are isodensity contours of the stellar disc, which rotates counter-clockwise.

to – and so robust against – variations in these assumptions. The main difference in the bulk radial changes that occur in these two extremes is that there is additional exchange of angular momentum in the case of the smooth disc, which is an anticipated effect of the bar formation process.

## 4 A TOY MODEL

One of the fundamental difficulties in relating dynamical analyses of pure N-body simulations to nature is the fact that initial (birth) radii of stars are not directly observable quantities. This problem is frequently circumvented by assuming that stars form following some kind of initial metallicity gradient, such that the metallicity of stars can be used as an observational tracer of their birth radius. We here assume that stars in the Milky Way are born with metallicities that generally follow a radius-dependent gradient, and we employ a simple toy model with metallicities 'painted' on to the stars in our N-body simulation. This toy model, though an over-simplification, allows us to explore the potential signatures and trends of the bulk radial changes in metallicity space.

In our toy model, we assign metallicities to stars as a function of their (cylindrical, galactocentric) initial radius in our simulation, following the functional form in Di Matteo et al. (2013):

$$Z(R) = Z_0 10^{\gamma R_0} \qquad (2)$$

where $Z_0$ is three times the solar value, $R_0$ is the initial ('birth') radius at $t = t_0$, and $\gamma$ is the metallicity gradient. We set $\gamma = -0.21$ dex per $R_d$ (i.e. if adopting a Milky Way-like scaling given in Section 1, $\gamma = -0.07$ dex per kpc, generally matching present-day observations of younger red giant branch stars, see e.g. Anders et al. 2017). We adopt $\frac{Z_\odot}{X} = 0.0207$ (consistent with the PARSEC stellar evolutionary tracks, Bressan et al. 2012), such that the metallicity at the center of the galaxy is [M/H] $\sim 0.5$. As we are assuming that all stars have the same age, the resulting metallicity trends would be most readily comparable to those in a mono-age population across the disc.

The amplitude of the angle-dependent $\Delta R$ variations are dependent on the strength of the non-axisymmetric perturbation (e.g. the bar), while variations in metallicity space will have an additional dependence on the slope of the initial metallicity gradient (Di Matteo et al. 2013). We note that our results are not strongly dependent on the slope or functional form of the initial gradient – as long as there is some type of measurable radius-dependent initial gradient, there should exist a mapping between $\Delta R$ and metallicity variations. We illustrate this in Figure 6, which shows the angle-dependent median metallicity variations ($\Delta$[M/H], the median metallicity value of the stars within the same 4° azimuth bins used in Section 2.3) if different values of the gradient ($\gamma$) and dispersion ($\sigma$) are adopted. We incorporate metallicity dispersion by adding to the metallicity of each star a random value drawn from a Gaussian distribution centred on zero with width $\sigma$. For clarity, only the metallicity variations in the annuli centred on 1, 2, and 4 $R_d$ at $\Delta t = 1.5$ T are shown.

In this toy model, the radial rearrangement occurring in the disc results in angle-dependent variations in the median metallicity at a given annulus. Metal-rich regions in an annulus (i.e. peaks in $\Delta R$) are aligned with the angle of the bar to approximately $R_{bar}$, and are increasingly less well-aligned at larger radii (in agreement with Di Matteo et al. 2013). The inner zone, where orbits have evolved inwards, is more metal-poor on average than its initial metallicity (at $t = t_0$), while the outer zone, where orbits have evolved outwards, is more metal-rich.

### 4.0.1 Caveats

The presence of a present-day radial metallicity gradient for stars of a given age does not necessarily imply that those stars were born with a metallicity distribution that followed that gradient. Indeed, as discussed in Schönrich & McMillan (2017), the present-day radial gradient is determined by the gradient of the star-forming interstellar medium, the spatial dependence of star formation, and the radial rearrangement of stars in the disc. It is also likely that a galactic bar would impact star formation and gas enrichment, and these effects may not be fully captured with a smooth radius-dependent initial metallicity gradient. It is perhaps for these reasons that sinusoidal angle-dependent metallicity variations, while common, are not necessarily ubiquitous in barred simulations (for example, the N-body smoothed particle hydrodynamics simulation presented in Kubryk et al. 2013 has angle-dependent metallicity variations dissimilar to those discussed here, see their Figure 13). The N-body simulation that we use for this analysis lacks gas, star formation, and chemical evolution, and thus this discussion is necessarily over-simplified. We leave further investigation of these important effects to future works.

It is prudent to investigate alternative, metallicity-independent tracers, especially given the uncertainties described above and the need for an observable quantity to link dynamical analyses (such as this one) to nature. For example, it may be possible to identify bulk radial changes in the disc via stellar age distributions if the galaxy had a radius-dependent star formation history.

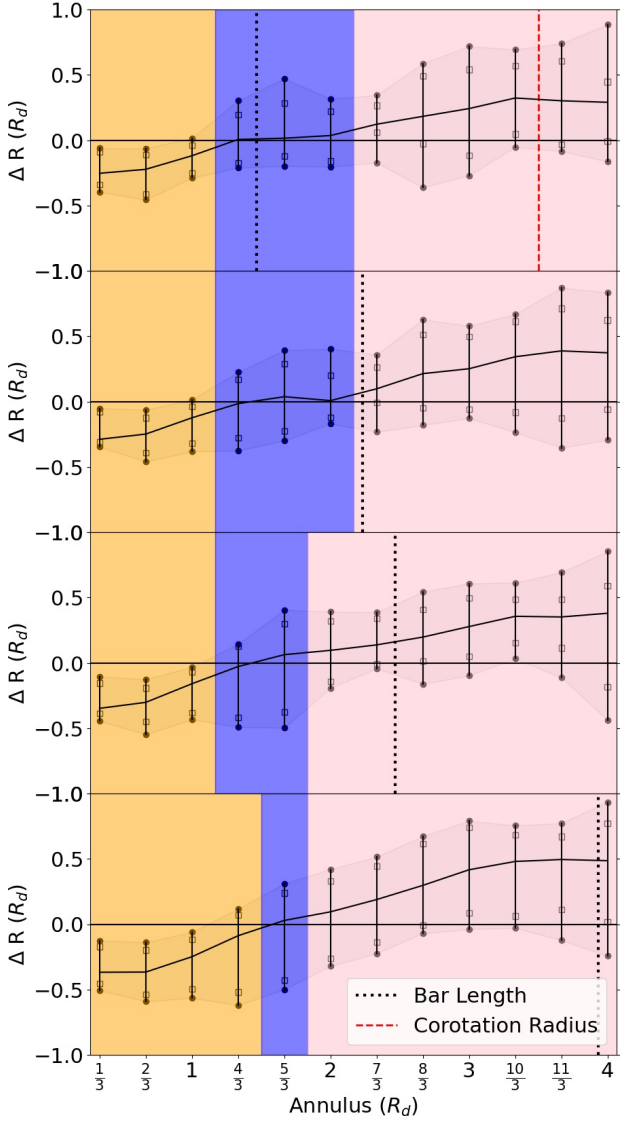

**Figure 5.** The maximum and minimum $\Delta R$ (solid black circles) for stars in each of the twelve annuli at the four time intervals considered in this analysis (rows correspond to $\Delta t = 0.5, 1.0, 1.5, 2.0$ T from top to bottom) alongside the bar length and approximate corotation radius at that time (black dotted and red dashed lines, respectively, note that at most times corotation is beyond $4R_d$). The tenth and ninetieth quantiles of $\Delta R$ are shown as black open squares, and the $\langle \delta R \rangle_\phi$ in each annulus is plotted as the solid black line. The inner zone, defined to be where $\langle \delta R \rangle_\phi \leq -\frac{1}{12}R_d$, is shaded in orange. The intermediate zone, where $-\frac{1}{12}R_d < \langle \delta R \rangle_\phi < \frac{1}{12}R_d$, is shaded in blue, and the outer zone, where $\frac{1}{12}R_d \geq \langle \delta R \rangle_\phi$, is shaded in pink. For reference, $R_{\text{bar}_0}$ is $\sim 1.4R_d$, respectively.

### 4.1 Observational Signatures in the Milky Way

Our model, while simplistic, allows us to consider how dynamics may be connected to observable trends in the Milky Way. The Milky Way bar is rather short ($R_{bar} \sim 5$ kpc, e.g. Wegg et al. 2015) and the bar length in our simulation is thus most comparable to the Milky Way at early times ($\Delta T \sim 0.5$). Corotation with the bar in the Milky Way is likely around $R_{CR} \sim 5$ kpc (see e.g. the recent analysis of Gaia Collaboration et al. 2022b) and there is additional spiral structure that we do not explicitly investigate in our simulation, further complicating direct comparison. Our simulation is not meant to reproduce the Milky Way, but is instead a laboratory within which to explore the impact of a galactic bar. We expect many of the trends described in this analysis to be generic for bar-dominated systems. For example, azimuth-angle dependent trends occurring with a bulk inwards evolution in the inner disk should be present in galaxies with a growing bar. Similarly, there should be peaks in $\Delta R$ aligned with the bar out to approximately the end of the bar. We refrain from making detailed predictions and comparisons beyond these generic trends.

Analyses of relatively sparse tracers (such as classical Cepheids) have revealed hints of angle-dependent stellar metallicity variations in the Milky Way disc (see e.g. Davies et al. 2009, Genovali et al. 2014, Kovtyukh et al. 2022). It is difficult to discern whether the variations seen in the sparse tracers are aligned with any given Galactic structure. Recent analyses using larger samples have further probed these variations and their possible alignments with Galactic structure (Gaia Collaboration et al. 2022a, Poggio et al. 2022, Hawkins 2023). The amplitude and morphology of the azimuthal variations in these larger samples is dependent on the tracer stellar population. Samples of giant stars within a few kiloparsec of the Sun show azimuthal metallicity variations at the level of $\sim 0.1$ dex, with the metal-rich regions aligned with the known spiral arm over-densities (Hawkins 2023, see also Poggio et al. 2022). However, this pattern is dependent on the effective temperature (and thus age) of the tracer population: Poggio et al. (2022) find that the alignment of metallicity variations with the spiral over-densities is seen most strongly in the hottest young stars ($\leq 100 - 300$ Myr) and is nearly non-existent in the cooler, relatively older stars. Similarly, Hawkins (2023) find that their sample of LAMOST OBAF stars has a different azimuthal pattern than the giant stars.

Indeed, kinematically cooler (i.e. younger) components are predicted to react more strongly to perturbations than kinematically hotter (i.e. older) components (e.g. Solway et al. 2012, Vera-Ciro et al. 2014, Vera-Ciro et al. 2016, Daniel & Wyse 2018). In discs with *vertical* initial metallicity gradients, this differential response can cause cause azimuthal metallicity variations (Khoperskov et al. 2018). However, given the very young ages of the hottest stars, the observed azimuthal variations in this population may be more indicative of recent star formation and enrichment trends.

Other stellar tracers have azimuthal metallicity variations of similar amplitude that do not appear to be aligned with spiral structure. The sample of OBAF stars shown in Hawkins (2023) and some samples of giant stars (dependent on the properties of the sample, see the 'RGB' sample in Gaia Collaboration et al. 2022a and the cooler giant sample in Poggio et al. 2022) have metallicity distributions that are asymmetric in the X-Y plane, with more metal-rich stars ahead of the Sun. Gaia Collaboration et al. (2022a) note that it is plausible that this concentration of metal-rich stars could be aligned with the bar, which generally matches the trends predicted by our simple toy model. Those authors further find that their RGB sample has lower metallicity than what would be predicted by a smooth radius-dependent initial metallicity gradient in the inner disc, as is also predicted by our toy model. Finally, we note that some analyses that target the inner disk with red clump and/or red branch giant stars find that the Galactic bar is more metal-rich than the surrounding disk (e.g. Wegg et al. 2019, Queiroz et al. 2021, Lian et al. 2021), matching our model predictions, while others do not (e.g. Bovy et al. 2019, Eilers et al. 2022). However, for the reasons given above, the utility of our toy model in these comparisons is limited. Additional analyses of the observational data (particularly those employing a larger range of $\phi$ angles and incorporating age information) and further simulations are required to investigate the plausible origins of these observed trends.

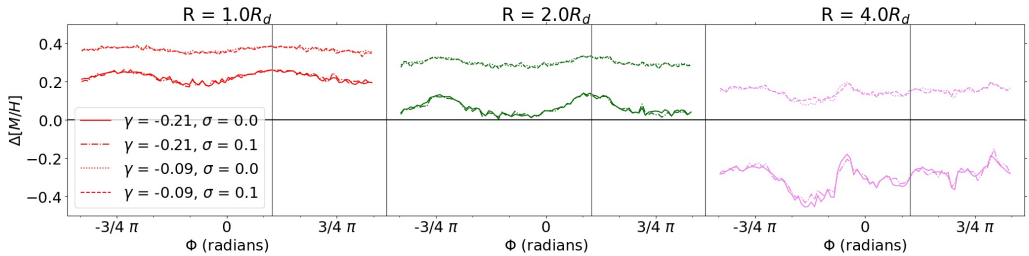

**Figure 6.** The angle-dependent median metallicity variations ($\Delta[\text{M/H}]$) in the annuli centred on 1, 2, and 3 $R_d$ at $\Delta t = 1.5$ T if different metallicity gradients and dispersions (line styles indicated in legend) are adopted. If employing a scaling such that $R_d = 3$ kpc, then the gradients of $\gamma = -0.21$ and $\gamma = -0.09$ dex per $R_d$ are equivalent to $-0.07$ and $-0.03$ dex per kpc, respectively.

### 4.2 Transforming Metallicity Information into Information on Bulk Radial Changes

There have been a number of analyses in the literature that aim to determine birth radii of individual stars in the Milky Way (see e.g. Minchev et al. 2018, Feltzing et al. 2020, Lu et al. 2022a). These analyses generally estimate the age for a given star and identify its likely birth radius by comparing its metallicity to a model of the initial metallicity distribution in the disc at the time of the star's birth. A similar analysis could be done to constrain the *bulk* radial changes that occurred in the disc. Indeed, as initial metallicity gradients are inherently information about *average* (or median) trends, we argue that it is more appropriate to use such a procedure to estimate the trends and amplitude of bulk radial changes than to determine birth radii for individual stars. Further, dispersion in the initial metallicity distribution could cause there to be significant uncertainties for individual birth radii (see e.g. Lu et al. 2022b), but it should have minimal impact on the measured mean or median metallicity for a given radius and azimuthal bin (see Figure 6).

#### 4.2.1 Learning About the Initial Gradient

If there are multiple episodes of star formation after the bar formed, the bulk radial rearrangement trends in a given annulus should be broadly similar in each of the resulting populations after a few orbital periods. The differential response of cooler and hotter components (as discussed above, e.g. Khoperskov et al. 2018) could cause the amplitude of the $\Delta R$ (or metallicity) signal to differ between populations of different ages, but should not affect the over-all behavior of stars in the three zones in the disc. It may be possible to use this idea to constrain the time evolution of the initial metallicity gradient. For example, one could divide the stars in the disc into different age bins and then estimate the $\Delta R$ from the median metallicity information, as outlined above, with some assumed model for the initial metallicity gradient. Large disagreement in resulting bulk trends from different age bins at the same annulus could indicate that the adopted model of the initial metallicity gradient is inappropriate. It could further be required that the innermost annuli exhibit net inwards evolution (i.e. negative $\Delta R$) if it is known that the bar is evolving.

The feasibility of providing such constraints in nature depends on the details of how star formation and chemical enrichment are carried out in barred galaxies, as well as how accretion of both stars and gas over time alter the chemical composition and gravitational potential. Additional investigations with chemo-dynamical (and possibly extremely high resolution cosmological) simulations are required to further test the predictions given here, which we leave to future studies.

### 4.3 Future Prospects

Spectroscopic observations focused on younger-to-intermediate age stars that cover a wide range of $\phi$ angles in the inner $\sim 5$ kpc of the Milky Way would be optimal for detecting bar-driven trends such as those predicted by our toy model. Younger stars should have experienced fewer total large-scale changes to the galactic potential compared to older stars that will have experienced the formation of the bar and potentially the most recent major merger (i.e. *Gaia* Sausage-Enceladus, Belokurov et al. 2018, Helmi et al. 2018, which would both infuse fresh gas for star formation and change orbits via altering the gravitational potential). The observed stars should be young enough to have been born after the bar formed, but old enough to have completed a few orbits. Beyond defining the sample, age estimates would be critical to the analysis as the mapping between $\Delta R$ and metallicity is most accurate for stars of approximately the same age.

Compared to the Solar neighborhood, observations in the inner disc can more easily obtain the $\phi$ coverage needed to characterise any angle-dependent metallicity variations: mapping just a quarter of the predicted sinusoidal $\phi$ versus metallicity curve requires a $\phi$ coverage of $\sim 90°$. The inner disc is also where the bar-driven, angle-dependent $\Delta R$ (and thus metallicity) trends should be most coherent and aligned with the bar (see above, also note the Milky Way has a bar length of $\sim 5$ kpc, Wegg et al. 2015). The currently available spectroscopic data are rather patchy in the interior regions of the Galaxy, but the upcoming SDSS-V, 4MOST, and MOONS spectroscopic surveys will each map the inner disc (see e.g. Kollmeier et al. 2017, Chiappini et al. 2019, Gonzalez et al. 2020) and provide relevant data.

Looking beyond the Milky Way, we note that it may soon be possible to detect azimuth angle-dependent metallicity variations in the resolved stellar populations of M31 using spectra of individual red giant branch stars. Given the known non-axisymmetric features (e.g. a bar, Athanassoula & Beaton 2006) it is likely that M31 is host to azimuth angle-dependent trends in $\Delta R$, which may or may not be similar to those in the Milky Way. Indeed, analyses of photometric metallicities (e.g. Gregersen et al. 2015) and integral field unit (IFU) data (e.g. Saglia et al. 2018, Gajda et al. 2021) have indicated that there may be a metal-rich region aligned with the bar. The first wide-field, massively multiplexed spectroscopic surveys that are capable of measuring metallicities of individual red giant stars in M31 are coming online, such as PFS and DESI (see e.g. Takada et al. 2014,

Dey et al. 2023), which can provide the data to enable more detailed exploration of azimuthal metallicity variations in the disc of M31.

## 5 COMPARISON TO MORE DISTANT GALAXIES

Our toy model can also be compared to IFU observations of more distant external galaxies. The benefit of comparing to external galaxies is that spectroscopic observations of face-on or near face-on barred galaxies can map azimuthal variations over the full two $\pi$ range of angles. This would, in principle, enable the creation of plots similar to Figure 3. This benefit comes at the cost of integrating over all of the light in a given area such that the resulting spectrum is a combination of all of the gas and stellar populations in that region. Care should also be taken when comparing observations to our simulation since star particles are most similar to a single age stellar population.

Spectroscopic analyses of external galaxies have found azimuthal variations in stellar age and abundances (e.g. Sánchez-Blázquez et al. 2011, Fraser-McKelvie et al. 2019, Neumann et al. 2020 and references therein), as well as in ages abundances (e.g. Sánchez-Menguiano et al. 2016, Ho et al. 2017, Kreckel et al. 2019 - note we ignore gas throughout this analysis, and we thus make no comparison to these or similar results). Some of the stellar population trends in external barred galaxies are generally in agreement with our findings. For example, analyses have found that bars are generally slightly older and more metal-rich than the area of the disc immediately surrounding the bar (see e.g. Sánchez-Blázquez et al. 2011, Neumann et al. 2020). If the disc formed inside-out with a negative age and metallicity gradient, these observations are aligned with our predictions (i.e. peaks in $\Delta R$ = peaks in metallicity = peaks in age distribution, with peaks generally aligned with the bar major axis out to the end of the bar).

If stars are born following a radius-dependent initial metallicity gradient, the effect of the orbital evolution in the three zones would cause the azimuth angle-averaged radial metallicity gradient in the inner and outer regions to have a shallower slope than the initial gradient (and gradient in the intermediate zone) for a stellar population of a given age. This could lead to barred galaxies appearing to have a shallower gradient overall than non-barred galaxies, depending on the spatial resolution and radii over which the gradient is measured. For example, high spatial resolution data over the full disk of a barred galaxy may be able to identify breakpoints in the gradient where the slope becomes more shallow in the inner and outer disk, whereas that detail may be averaged out to appear as a generally shallower gradient in data of lower spatial resolution and/or smaller areal coverage. There does not appear to be consensus in the literature as to whether or not barred galaxies are observed to have shallower metallicity gradients than unbarred galaxies (see e.g. Seidel et al. 2016 and references therein for discussion on the disagreement in the literature), which may suggest that differences in techniques and spatial resolution/coverage are important. These disagreements indicate a need for future observational analyses.

## 6 DISCUSSION

The $\Delta R$ and $\langle \delta R \rangle_\phi$ trends discussed in this analysis have revealed that the response of the disc to a bar is both radius- and azimuth angle- dependent. Further, the effects of dynamic response and secular evolution are coupled. While we compute $\Delta R$ and $\langle \delta R \rangle_\phi$ with $t_0$ defined to be post bar-formation, the results highlighted here remain true even if the initial time instead corresponds to the start of the simulation ($t = 0$), before structure forms. The main difference between these two scenarios is that when $t = 0$, there is additional angular momentum transfer associated with the formation of the bar that causes $\Delta R$ and $\langle \delta R \rangle_\phi$ to tend towards higher positive values (increase in radius). The three radial zones shift accordingly.

The radial extent of the three zones would likely differ for bars of different speeds and lengths. Indeed, in Figure 5, it is evident that the radial extents of the zones change over time in our simulation as the bar lengthens and slows. The intermediate zone is largest at early times, when the bar is relatively fast and short, and it shrinks as the bar slows. It is possible that this holds generally, and that galaxies with faster bars have larger intermediate zones. However, as this is only one realization of a barred galaxy, additional analyses of the extent of the three zones in other galaxies is needed.

The bulk trends in the inner disc, i.e. inwards evolution in concert with angle-dependent trends that are aligned with the bar, indicate that stars can experience both an increase in eccentricity and a change in angular momentum (see also Ghosh et al. 2022 for further discussion). One can see how this naturally arises within the region most strongly dominated by the bar. Here, the $\phi$ location of the (radial) apoapses of stars tend to align with the bar major axis. As orbits become more eccentric, they create peaks of $\Delta R$ that are then aligned with the bar major axis, while troughs will be aligned with the minor axis. These peaks and troughs occur in addition to whatever bulk exchange of angular momentum occurs. Analyses that focus solely on stars that change angular momentum but do not become eccentric, and vice-versa, thus paint an incomplete picture of the evolution within the disc. Similarly, analyses focusing only on the angle-dependent trends may miss the broader radius-dependent evolution in the disc. Below, we comment on two additional implications of our results.

### 6.1 Morphology of Bulk Radial Changes in Context

Galactic bars are not the only possible source of angle-dependent $\Delta R$ variations – spiral arms and the gravitational perturbations from a merging satellite could also cause such signals. As discussed above, bar-driven, angle-dependent $\Delta R$ variations will be aligned with the bar within the bar-dominated region and will start to be offset from the bar angle at radii larger than the bar length, where trailing structure begins. Indeed, the degree of alignment of $\Delta R$ variations with the bar angle could help discern where the transition between bar and trailing structure occurs. Spiral-driven angle-dependent $\Delta R$ variations will align with the spiral (see e.g. metallicity variations in Khoperskov et al. 2018) such the phase of the $\phi$ versus $\Delta R$ curve will vary as a function of radius.

The $\Delta R$ variations induced by an orbiting satellite need not align with any internal structure, and should be strongest near the location of the pericenter passage. Carr et al. (2022) show for a Milk Way-like galaxy experiencing a Sagittarius-like merger, the variations of the change of angular momentum and metallicity in the X-Y plane will look somewhat like a quadrupole but will not necessarily be symmetric in distribution or strength. The differing behavior of the $\phi$-dependent trends as a function of radius resulting from these three different origins provide hope that it may be possible to disentangle the cause of observed angle-dependent trends if given sufficient spatial information. Note that in the outer disc within a limited spatial range the three signals discussed here may appear quite similar. As such, special care should be taken in interpreting the origins of any $\phi$ versus $\Delta R$ trends at large $R$.

**6.2 Implications for Semi-Analytic Modelling**

The results presented here should be included in future semi-analytic modelling analyses of the radial evolution of orbits (see e.g. Schönrich & Binney 2009, Brunetti et al. 2011, Frankel et al. 2018 for recent examples of such analyses). Such analyses typically model the radial changes of stars in the disc as a diffusion process with some dependence on age. However, a diffusion process alone does not describe the bulk radial changes driven by a galactic bar. As shown in this analysis, it is crucial to include a radial and azimuthal dependence on the distance that stars travel within the disc of a barred galaxy. Angle-dependence in semi-analytic modelling is likely also important for spiral galaxies with weak or non-existent bars (see e.g. Grand et al. 2016, Khoperskov et al. 2018).

**7 CONCLUSION**

We investigate the radial re-distribution of stars that occurs in an N-body simulation of a disc galaxy after a bar has formed. We summarise the conclusions of this analysis as follows:

- The bar induces azimuth angle-dependent and radius-dependent trends in the median distance that stars have travelled to enter a given annulus. The azimuthal trends are aligned with the bar out to approximately the end of the bar and become offset at larger radii. Angle dependent trends are present throughout the disc.
- The radius-dependent trends divide the disc into three zones. On average, stars currently in the inner zone (where $\langle \delta R \rangle_\phi$ is less than zero, here $R \lesssim R_d$, $R \lesssim R_{\mathrm{bar}_0}$) originated at larger radii and their orbits have evolved inwards. Stars currently in the intermediate zone (where $\langle \delta R \rangle_\phi$ is near zero, here $R \gtrsim R_d$, $R \gtrsim R_{\mathrm{bar}_0}$) could have originated at larger or smaller radii, and there is no net radial evolution (i.e. $\Delta R$ is roughly symmetric about zero). Finally, stars in the outer zone (where $\langle \delta R \rangle_\phi$ is greater than zero, here $R > R_{\mathrm{bar}_0}$) on average originated at smaller radii and their orbits evolved outwards.
- We assign metallicities to the star particles in our simulation following a radial metallicity gradient and explore the metallicity trends in this toy model. The radial rearrangement in the disc causes the inner and outer zones become more metal-poor and metal-rich, respectively, than their initial metallicities. Peaks of $\Delta R$ in a given annulus correspond to metal-rich regions in this model.
- We discuss recent observational evidence for angle-dependent metallicity variations in the Milky Way and external galaxies, and find that some of these trends are plausibly related to the bar-driven trends predicted here. We comment on the possibility of using observed angle-dependent median metallicity trends to learn about the initial metallicity gradient and the bulk radial changes that have occurred in the disc.

Future analyses of the angle- and radius-dependent trends in barred galaxies are necessary to further explore the ideas presented here. For example, it would be prudent to test the predictions given in Section 4 with simulations of barred galaxies that include gas, star formation, and galactic mergers. Observation-based analyses also hold exciting promise as new spectroscopic surveys enable wider and more dense coverage of the Milky Way disc and beyond.

**DATA AVAILABILITY**

The simulation used in this analysis will be made available upon reasonable request. The following codes were used in the analysis and production of this manuscript: EXP (Weinberg 1999; Petersen et al. 2022), Matplotlib (Hunter 2007), numpy (Harris et al. 2020), pandas (Wes McKinney 2010), and scipy (Virtanen et al. 2020).

**ACKNOWLEDGMENTS**

The authors thank the Flatiron Institute's Center for Computational Astrophysics and the Big Apple Dynamics school for making this work possible. CF would like to thank Keith Hawkins, Chris Carr, Mike Petersen, Jason Hunt, and Rosemary F.G. Wyse for insightful conversation. CF acknowledges support by the NASA FINESST grant and support through the generosity of Eric and Wendy Schmidt, by recommendation of the Schmidt Futures program. Contributions by RLM to this material are supported by the Wisconsin Space Grant Consortium under NASA Award No. 80NSSC20M0123, the Ruth Dickie Endowment of the UW-Madison Beta Chapter of SDE/GWIS, and the National Science Foundation Graduate Research Fellowship under Grant No. 2137424. KJD gratefully acknowledges the Flatiron Institute's Center for Computational Astrophysics for hosting her during the course of this work.

## APPENDIX A: SUPPLEMENTARY PLOTS

We include here additional plots that aid in the characterisation of our simulation. We present the rotation curve of the galaxy at the beginning of the simulation in the left panel in Figure A1, where the disk, halo, and combined rotation curves are each shown with a different line style and colour-coding. In Figure A2, we show the angle-averaged radial distribution of the stars between zero and 10 $R_d$ at the start of the simulation (IC) and at $\Delta t = 0$, 0.5, 1.0, 1.5, and 2.0 T, along with the radial location of the peak of the distribution at each time. The redistribution of stars throughout the disk over time is evident. Finally, in Figure A3, we present the buckling amplitude and asymmetry parameters as a function of time, computed following

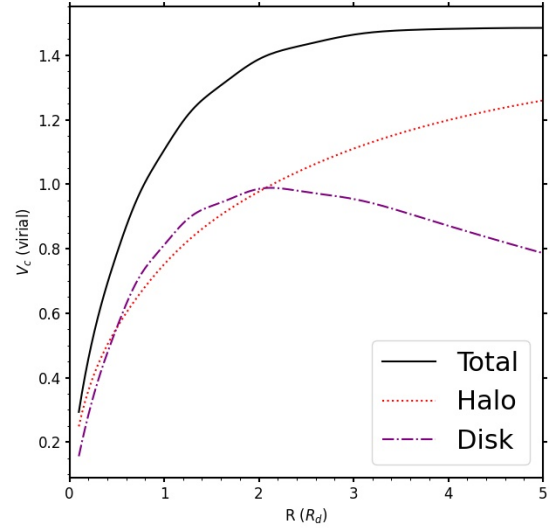

**Figure A1.** The initial rotation curve for the disk (purple dot-dashed line) and halo (red dotted line) components, alongside the total (black solid line) rotation curve.

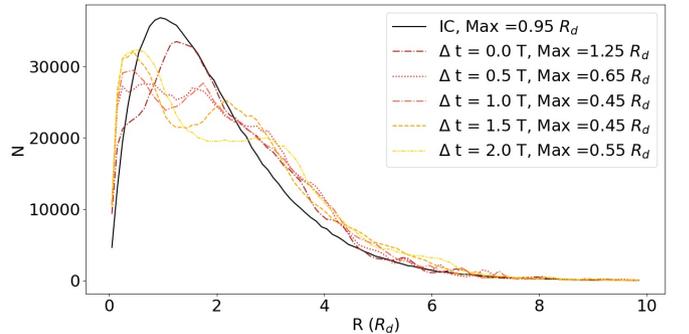

**Figure A2.** The radial distribution of stars between zero and 10 $R_d$ at the start of the simulation (IC) and $\Delta t = 0.0, 0.5, 1.0, 1.5,$ and $2.0$ T. Each time is indicated by the line style and colour-coding given in the legend, along with the approximate radius where the density distribution is maximum. The rearrangement of stars (and angular momentum) can easily be seen from the change in the distribution.

(e.g.) Baba et al. (2022) and Smirnov & Sotnikova (2018), respectively. As in Section 2, we compute these metrics over all of the stars within an annulus between 0.25 and 0.50 $R_d$ in the disk. We smooth the resulting plots with a Savitzky–Golay filter (as implemented in Scipy, Virtanen et al. 2020) with a window of 51 and a polynomial order of three. The buckling amplitude for m = 1 and m = 2 (orange solid and red dotted lines, respectively) is never comparable to the vertical scale height, indicating that no buckling occurs. Similarly, the asymmetry parameter (purple solid line) is always ≲ 0.1 once the bar forms.

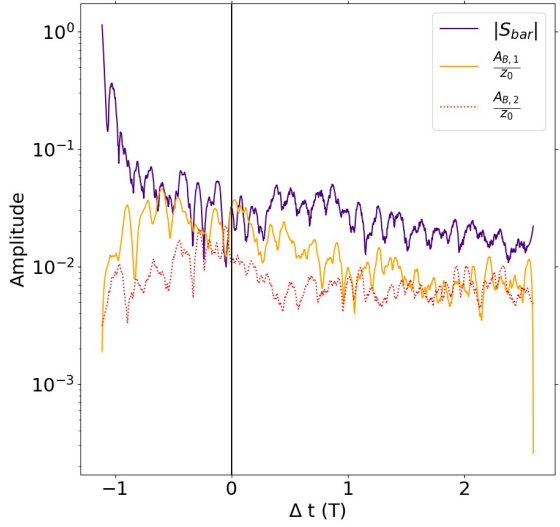

**Figure A3.** The buckling amplitude for m = 1 (orange solid line) and m = 2 (red dotted line) alongside the absolute value of the asymmetry parameter ($|S_{bar}|$, purple). Each metric is computed over all of the stars within a radial annulus from 0.25 to 0.5 $R_d$ in the disk, and is then smoothed with a Savitzky–Golay filter. For clarity, we show the y-axis on a log scale.